\documentclass[sort,sort&compress,a4paper]{elsarticle} % Elsevier style LaTeX class

\usepackage{setspace}%
\newcommand{\gev}{\ensuremath{\;\mathrm{GeV}}} %
\newcommand{\mev}{\ensuremath{\;\mathrm{MeV}}} %
\newcommand{\kev}{\ensuremath{\;\mathrm{keV}}} %
\newcommand{\mpc}{\;\mathrm{Mpc}} % Mpc
\newcommand{\dm}{{\textsc{dm}}} % dark matter
\newcommand{\numsm}{$\nu$MSM\xspace} % nuMSM

\usepackage{units} % Command $\unit[10]{GeV}$ prints correctly 10 GeV. Command
\IfFileExists{srcltx.sty}{\usepackage[active]{srcltx}}

\usepackage{geometry,fullpage}
\usepackage{paralist,color,xspace,eurosym,amsmath,amssymb,amsfonts,bm}
\usepackage{array,hhline,graphicx,subfig}

\begin{document}

\title{Next decade of sterile neutrino studies} %

\author{Alexey Boyarsky$^{1,2,3}$,  Dmytro~Iakubovskyi$^{1,3}$, Oleg Ruchayskiy$^{4,2}$\\
  $^1${\small Instituut-Lorentz for Theoretical Physics, Universiteit Leiden,}\\
  {\small Niels Bohrweg 2, Leiden, The Netherlands}\\
  $^2${\small Ecole Polytechnique F\'ed\'erale de Lausanne,}\\
  {\small FSB/ITP/LPPC, BSP
    720, CH-1015, Lausanne, Switzerland}\\
  $^3${\small Bogolyubov Institute of Theoretical Physics, Kyiv, Ukraine}\\
  $^4${\small CERN Physics Department, Theory Division,}\\
  {\small CH-1211 Geneva 23, Switzerland} %
}\date{}

\begin{abstract}
We review the status of sterile neutrino dark matter and discuss astrophysical and cosmological bounds 
on its properties as well as future prospects for its experimental searches. We argue that 
if sterile neutrinos are the dominant fraction of dark matter, detecting an astrophysical signal from 
their decay (the so-called `indirect detection') may be the only way to identify these particles 
experimentally. However, it may be possible to check the dark matter origin of the observed signal 
unambiguously using its characteristic properties and/or using synergy with accelerator experiments, 
searching for other sterile neutrinos, responsible for neutrino flavor oscillations. 
We argue that to fully explore this possibility a dedicated cosmic mission -- an X-ray spectrometer -- is needed.
\end{abstract}

\maketitle

\section{Dark matter problem and particle physics}
\label{sec:dark-matt-cand}

The nature of dark matter (DM) is among the most intriguing questions of modern physics. 
There is a body of strong and convincing evidence of its existence. 
Indeed, numerous independent tracers of gravitational potential 
(observations of the motion of stars in galaxies and galaxies in clusters; 
emissions from hot ionized gas in galaxy groups and clusters; 
21~cm line in galaxies; both weak and strong gravitational lensing measurements) 
demonstrate that the dynamics of galaxies and galaxy clusters cannot be explained 
by the Newtonian potential created by visible matter only. 
Moreover, cosmological data (analysis of the cosmic microwave background anisotropies 
and of the statistics of galaxy number counts) shows that the cosmic large scale structure 
started to develop much before decoupling of photons at recombination of hydrogen in the 
early Universe and, therefore, much before ordinary matter could start clustering. 
This body of evidence points at the existence of a new substance, 
universally distributed in objects of all scales and providing a contribution to the 
total energy density of the Universe at the level of about 25\%. 
Various attempts to explain this phenomenon by the presence of macroscopic compact objects 
(such as, for example, old stars) or by modifications of the laws of gravity (or of dynamics) 
failed to provide a consistent description of all the above phenomena.
The abundance of baryonic dark matter is strongly constrained by numerous microlensing experiments 
(in form of MAssive Compact Halo Objects in mass range from $\sim 10^{-7} M_\odot$ to $\sim 10~M_\odot$, 
see e.g.~\cite{Gates:95,Lasserre:00,Alcock:00}; for an overview see
\cite{Moniez:10} and references therein) and the results of Big Bang Nucleosynthesis~\cite{Dar:95}.
Attempts to explain dark matter by the existence of primordial
black holes have not been fully successful (see e.g.~\cite{Carr:05,Capela:12}). 
 Therefore, a microscopic origin of dark matter phenomenon (i.e.\ a new particle or particles) 
remains appealing hypothesis.

The only electrically neutral and long-lived particle in the Standard Model of particle physics (SM)
are neutrinos. 
As the experiments show that neutrinos have mass, they could play the role of dark matter particles. 
Neutrinos are involved in weak interactions that keep these particles in the early Universe in thermal 
equilibrium down to the temperatures of few MeV. At smaller temperatures, 
the interaction rate of weak reactions drops below the expansion rate of the Universe and neutrinos 
``freeze out'' from the equilibrium. 
Therefore, a background of relic neutrinos was created just before primordial nucleosynthesis. 
As interaction strength and, therefore, decoupling temperature and concentration of these particles are known, 
their present day density is fully defined by the sum of the masses for all neutrino flavors. 
To constitute the whole DM this mass should be about 11.5~eV (see e.g.~\cite{Lesgourgues:06}). 
Clearly, this mass is in conflict with the existing experimental bounds: measurements of the electron 
spectrum of $\beta$-decay put the combination of neutrino masses below 2~eV~\cite{PDG-12} while from the 
cosmological data one can infer an upper bound of the sum of neutrino masses is 
$\sum m_i \lesssim 1$~eV~\cite{Lesgourgues:2012uu}.
The fact that neutrinos could not constitute 100\% of DM follows also from the study of 
phase space density of DM-dominated objects that should not exceed the density of degenerate Fermi gas: 
fermionic particles could play the role of DM in dwarf galaxies only if their mass is above few hundreds of eV 
(the so-called `Tremaine--Gunn bound'~\cite{Tremaine:79}, for review see \cite{Boyarsky:08a} and references 
therein) and in galaxies if their mass is above few tens of eV. 
Moreover, as the mass of neutrinos is much smaller than their decoupling temperature, 
they decouple relativistic and become non-relativistic only in matter-dominated epoch 
(``hot dark matter''). For such a dark matter the history of structure formation would be very different and 
the Universe would look rather differently nowadays~\cite{Davis:85}. 
All these strong arguments prove convincingly that \emph{dominant fraction of dark matter}
cannot be made of the Standard Model neutrinos and therefore \emph{the Standard Model of
  elementary particles does not contain a viable DM candidate.} 
Therefore, the DM particle hypothesis necessarily implies an extension of the SM, see 
e.g.~\cite{Bertone:04,Bertone:10book,Taoso:07,Drees:12} for a review of DM particle candidates.

Phenomenologically little is known about the properties of DM particles. 
The mass of fermionic DM is limited from below by the `Tremaine--Gunn bound'\footnote{A much weaker bound, 
based on the Liouville theorem, can be applied for bosonic DM, see e.g.~\cite{Madsen:90,Madsen:91}.}.
They are not necessarily stable, but their lifetime
should significantly exceed the age of the Universe 
(see e.g.~\cite{Boyarsky:08b}); DM particles should have become non-relativistic sufficiently early in the 
radiation-dominated epoch (although a sub-dominant fraction might have remained relativistic much later). 

%\bigskip

A lot of attention has been devoted to the class of dark matter candidates called 
\emph{weakly interacting massive particles} (WIMPs) (see e.g.~\cite{Bertone:04,Feng:10} for review). 
These hypothetical particles generalize the neutrino DM~\cite{Lee:77}: 
they also interact with the SM sector with roughly electroweak strength, 
however their mass is large enough so that these particles become non-relativistic already at decoupling. 
In this case the present day density of such particles depends very weakly (logarithmically) 
on the mass of the particle as long as it is heavy enough. 
This ``universal'' density happens to be within the order of magnitude consistent with DM density 
(the so-called ``\emph{WIMP miracle}''). Due to their large mass and interaction strength, 
the lifetime of these particles would be extremely short and therefore some special symmetry 
has to be imposed in the model to ensure their stability.

The interest for this class of candidates is due to their potential relation to the electroweak 
symmetry breaking, which is being tested at the LHC in CERN. 
In many models trying to make the Standard Model ``natural'' like, for example, 
supersymmetric extensions of the Standard Model, there are particles that could play the role of 
WIMP dark matter candidates. The WIMP searches are important scientific goals of many experiments. 
Dozens of dedicated laboratory experiments are conducted to detect WIMPs in the Galaxy halo by testing 
their interaction with nucleons (\emph{direct detection experiments}) (see e.g.~\cite{Saab:2012th} 
and references therein). Searches for the annihilation products of these particles (\emph{indirect detection}) 
are performed by PAMELA, Fermi and other high-energy cosmic missions 
(see e.g. reviews~\cite{Lavalle:12,Bergstrom:12a}). No convincing signals has been observed so far 
in either ``direct'' or ``indirect'' searches.

Additionally, no hints of new physics at electroweak scale had turned up at the LHC or in any other experiments. 
This makes alternative approaches to the DM problem ever more viable.

\section{Sterile neutrino dark matter}
\label{sec:sterile-neutrino-dm}

\begin{figure}[!tp]
  \centering 
  \subfloat[Decay of sterile neutrino $N \to \nu_e \nu_\alpha \bar\nu_\alpha$
  through neutral current interactions. A virtual $\nu_e$ is created and the
  quadratic mixing (marked by symbol ``$\bm{\times}$'') is proportional to
  $\theta^2$.]{\label{fig:N23nu}\includegraphics[width=.3\textwidth]{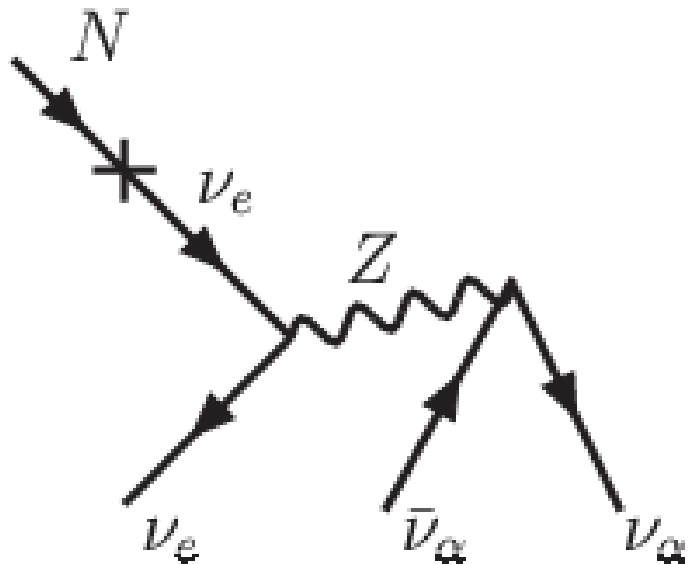}}
  ~~$\Longrightarrow$~~ \subfloat[At energies $E_N\ll M_Z$, the process in the
  left panel can be described by the Fermi-like interaction with the
  ``effective'' Fermi constant $\theta_e\times G_F$.]
  {\label{fig:eff-fermi}\includegraphics[width=.3\textwidth]{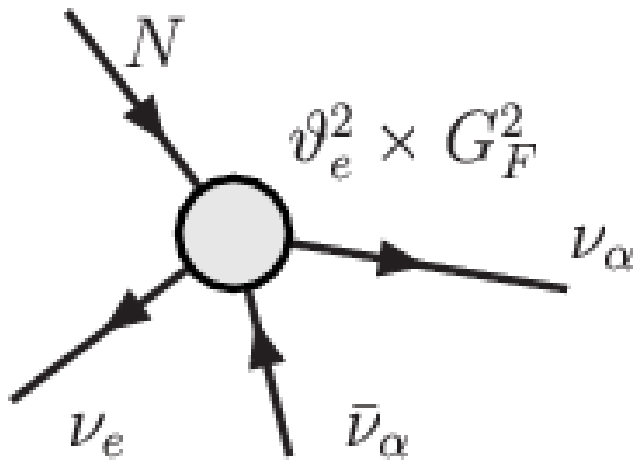}
  } ~~~~ \subfloat[Two-body decay of sterile neutrino. The energy of the
  photon is $E_\gamma = \frac 12 M_N$.]
  {\label{fig:decay}\includegraphics[width=.3\textwidth]{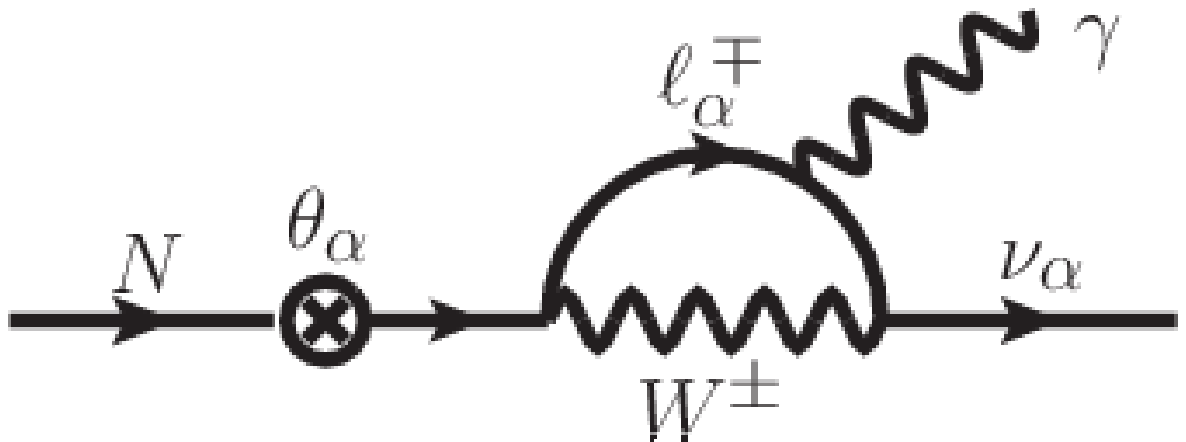}
  }
  \caption{Example of interactions of sterile neutrino: decay $N\to \nu_e
    \nu_\alpha \bar\nu_\alpha$ (panel~\ref{fig:N23nu}) and its effective
    Fermi-like description (panel~\ref{fig:eff-fermi}) and loop-mediated
    decay $N\to \gamma + \nu_\alpha$ (panel~\ref{fig:decay}).}
  \label{fig:fermi1}
\end{figure}

Another viable generalization of the neutrino DM idea is given by \emph{sterile neutrino dark matter}
scenario~\cite{Dodelson:93,Shi:98,Dolgov:00,Abazajian:01a,Abazajian:01b,Asaka:05a},
see~\cite{Boyarsky:09a,Kusenko:09a} for review. 
Sterile neutrino is a right-chiral counterpart of the left-chiral neutrinos of the SM 
(called `\emph{active}' neutrinos in this context). Adding these particles to the SM Lagrangian 
makes neutrinos massive and is therefore their existence provides a simple and natural explanation 
of the observed neutrino flavor oscillations. These particles are singlet leptons because they carry 
no charges with respect to the Standard Model gauge groups (hence the name), 
and therefore along with their Yukawa interaction with the active neutrinos (=`Dirac mass') 
they can have a Majorana mass term (see e.g.~\cite{Abazajian:2012ys} for details). 
They interact with the matter via creation of virtual active neutrino (quadratic mixing) 
and in this way they effectively participate in weak reactions (see e.g. Fig.~\ref{fig:N23nu} ). 
At energies much below the masses of the $W$ and $Z$-bosons, their interaction can be described by the analog 
of the Fermi theory with the Fermi coupling constant $G_F$ suppressed by the active-sterile neutrino 
mixing angle $\theta$ --- the ratio of their Dirac to Majorana masses (Fig.~\ref{fig:eff-fermi}):
\begin{equation}
  \label{eq:1}
  \theta_\alpha^2 = \sum_{\text{sterile N}}\left|\frac{m_{\text{Dirac},\,\alpha}}{M_\text{Majorana}}\right|^2
\end{equation}
(this mixing can be different for different flavours $\alpha$).

It was observed long ago that such particles can be produced in the Early Universe 
through mixing with active neutrinos~\cite{Dodelson:93} and have a correct relic density for any 
mass~\cite{Dodelson:93,Shi:98,Abazajian:01a,Asaka:06b,Asaka:06c,Shaposhnikov:08a,Laine:08a}.

The existence of sterile neutrinos is motivated by the \emph{observational phenomena beyond the Standard Model} 
(unlike WIMPs that are motivated first of all by the theoretical considerations of stability of the 
Higgs mass against quantum corrections that could require a fine-tuning of parameters of the model). 
Namely, sterile neutrinos would provide a simple and natural explanation of the \emph{neutrino flavour
  oscillations}~\cite{Minkowski:77,Ramond:79,Mohapatra:79,Yanagida:80}.
However, a \emph{single} sterile neutrino would be unable to explain the two observed mass splittings between 
Standard Model neutrinos -- at least two sterile neutrinos are needed for that. 
Moreover, should sterile neutrino play the role of DM, its mixing with active neutrinos would be too small 
to contribute significantly to the flavor oscillations -- its life time should be very large and, 
therefore, interaction strength should be too feeble ~\cite{Asaka:05a,Boyarsky:06a}. 
Therefore, in order to explain dark matter and neutrino mass (one for each SM flavor), 
the minimal model should contain 3 right-handed neutrinos~\cite{Asaka:05a}. 
In such a model, the lowest mass eigen-state of the active neutrinos will be (almost) zero 
and the sum of neutrino masses $\sum m_\nu \approx \kappa \sqrt{|\Delta m_\text{atm}^2|}$, 
where $\kappa = 1$ or $2$ for normal (inverted) hierarchy~\cite{Boyarsky:06a}. 
This is one of the predictions of such a model.

In spite of the fact that dark matter sterile neutrino plays essentially no role in the neutrino oscillations, 
the fact that 3 particles are needed to explain \emph{both} dark matter and neutrino oscillations is crucial. 
As we will see below, primordial properties of sterile neutrino dark matter are determined by 
two other sterile neutrinos.

If the masses of the two sterile neutrinos, responsible for neutrino oscillations, 
are below $\sim 2\gev$ (mass of $c$-quark), such particles can be searched with existing experimental 
techniques~\cite{Gorbunov:07a,gorbunov}, see Sec.~\ref{sec:accelerator-searches} below. 
This is a unique situation when one can directly test the nature of neutrino oscillations in 
`intensity frontier'~\cite{Intensity-frontier:11}  experiments. 
For masses above 2~GeV the searches become more difficult (see Sec.~\ref{sec:method} for details).

It turns out that in the region of masses between $100\mev$ and electroweak scale out-of equilibrium 
reactions with these two sterile neutrinos are capable of generating the observed matter–antimatter 
asymmetry of the Universe (baryogenesis)~\cite{Asaka:05b}. 
These observations motivated a lot of recent efforts for developing this model, called 
the \numsm\ --- \emph{Neutrino Minimal Standard Model} (see~\cite{Boyarsky:09a} for review). 
Therefore, finding these particles in intensity frontier experiments would provide an unparalleled 
possibility to test baryogenesis in laboratory. Moreover, if some particles are found in such experiments 
it will be possible not only to check whether they are responsible for baryogenesis or not, 
but also unambiguously predict the properties of sterile neutrino DM.

Because its interaction with the Standard Model particles is very feeble, sterile neutrino does not need 
to be stable. The decay channel for sterile neutrinos of all masses is to 3 (anti)neutrinos
(Fig.~\ref{fig:N23nu}).\footnote{For masses above $1\mev$ additional decay channels become kinematically possible.}  
However, the most characteristic feature of sterile neutrino DM is its ability to decay to photon 
and neutrino (with cosmologically long lifetime)~\cite{Pal:81,Abazajian:01b,Dolgov:00}, see
Fig.~\ref{fig:decay}. The emitted photon is almost mono-energetic 
(the width of the DM decay line is determined entirely by the motion of DM particles). 
Although the lifetime of the DM particles turns out to be \emph{much longer than the age of the Universe}, 
humongous amount of these particles around us implies that the combined emission may be sizable.

\emph{If dark matter is made of sterile neutrinos, detecting astrophysical signal from their decay 
(the ``indirect detection'') may be the only way to identify this particle experimentally. 
However, it may be possible to prove the dark matter origin of observed signal unambiguously 
using its characteristic properties.}

\bigskip

\emph{In summary}, one sees that three sterile with the masses below electroweak scale form a minimal
 testable model that provides a unified description of three major \emph{observational} problems 
``beyond-the-Standard-Model''~\cite{Asaka:05a,Asaka:05b,Boyarsky:09a,Canetti:12}:
\begin{compactenum}[1)]
  \item neutrino flavour oscillations;
  \item the absence of primordial anti-matter in the Universe;
  \item existence of dark matter.
\end{compactenum}

\subsection{Production of sterile neutrinos in the early Universe}

The interaction of active neutrinos with primordial plasma at temperatures above few MeV 
(see~\cite{Notzold:1987ik}) leads to a significant temperature suppression of active-sterile 
mixing~\cite{Barbieri:90} at temperatures above a few hundred MeV which therefore peaks roughly 
at~\cite{Dodelson:93,Dolgov:00,Asaka:06b}
\begin{equation}
  T_{peak} \sim 130\left(\frac{M_{N_\dm}}{\unit[1]{keV}}\right)^{1/3}~\mbox{MeV}.
  \label{peak}
\end{equation}
Sterile neutrinos DM are never in thermal equilibrium (see e.g.~\cite{Boyarsky:09a}) and their number density 
is significantly smaller 
than that of the active neutrinos (that is why they can account for the observed DM abundance without 
violating `Tremaine--Gunn bound'). In particular, the shape of the primordial momentum distribution 
of thus produced sterile neutrinos is roughly proportional to that of the active neutrinos~\cite{Dolgov:00}:
\begin{equation}
  \label{eq:19}
  f_{N_\dm}(t,p) = \frac{\chi}{e^{p/T_\nu(t)}+1}\;,
\end{equation}
where the normalization $\chi \sim \theta_\dm^2\ll 1$ and where $T_\nu(t)$ is
the temperature of the active neutrinos.\footnote{The true distribution of
  sterile neutrinos is in fact colder than that shown in Eq.~(\ref{eq:19}).
  Specifically, the maximum of $p^2 f_{N_1}(p)$ occurs at $p/T_\nu\approx
  1.5-1.8$ (depending on $M_{N_\dm}$), as compared with $p\approx 2.2 T_\nu$
  for the case shown in Eq.~(\ref{eq:19}) \cite{Asaka:06b,Asaka:06c}.}
Comparing the production temperatures Eq.~(\ref{peak}) of DM sterile neutrinos
with their masses shows that they are produced relativistically in the
radiation-dominated epoch. Indeed, for the primordial DM distribution of the
form~(\ref{eq:19}) one has $\langle p\rangle \sim T_{peak} \gtrsim M_{N_\dm}$
for $M_{N_\dm} \lesssim \unit[40]{GeV}$.  Relativistic particles stream out of
the overdense regions and erase primordial density fluctuations at scales
below the \emph{free-streaming horizon} (FSH) -- particles' horizon when they
becomes nonrelativistic (for a detailed discussion of characteristic scales
see e.g.~\cite{Boyarsky:08c} and references therein). This effect influences
the formation of structures. If DM particles decouple nonrelativistically
(\emph{cold} DM models, CDM) the structure formation occurs in a ``bottom-up''
 manner: specifically, smaller scale objects form first and then merge into the
larger ones~\cite{Peebles:80}. CDM models fit modern cosmological data well.
In the case of particles, produced relativistically and \emph{remaining
  relativistic} into the matter-dominated epoch (i.e. \emph{hot} DM, HDM), the
structure formation goes in a ``top-down'' fashion~\cite{Zeldovich:70}, where
the first structures to collapse have sizes comparable to the Hubble
size~\cite{Bisnovatyi:80,Bond:80,Doroshkevich:81}. The HDM scenarios
contradict large-scale structure (LSS) observations~\cite{Davis:85}.  Sterile
neutrino DM that is produced relativistic and is then redshifted to
nonrelativistic velocities in the radiation-dominated epoch is an
intermediate, \emph{warm dark matter} (WDM)
candidate~\cite{Bode:00,Abazajian:01a,Dolgov:00}. Structure formation in WDM
models is similar to that in CDM models at distances above the free streaming
scale. Below this scale density fluctuations are suppressed, compared with the
CDM case. The free-streaming scale can be estimated
as~\cite{Bond:80} \begin{equation}
  \label{eq:5} \lambda_\textsc{fs}^{co} \sim \unit[1]{Mpc}
  \left(\frac{\mathrm{keV}}{M_{N_\dm}}\right)\frac{\langle p_N\rangle}{\langle
    p_\nu\rangle}\;.  \end{equation} where $1\mpc$ is the (comoving) horizon
at the time when momentum of active neutrinos $\langle p_\nu \rangle \sim
1\kev$. If the spectrum of sterile neutrinos is nonthermal, then the
moment of non-relativistic transition and $\lambda_\textsc{fs}^{co}$ is shifted by $\langle
p_{N}\rangle/\langle p_\nu\rangle$.

This mechanism specifies a \emph{minimal} amount of sterile neutrinos that will be produced for given $M_1$ and $\theta_1$. 
The requirement that 100\% of DM be produced via such mixing places an \emph{upper bound} on the mixing angle 
$\theta_1$ for a given mass. This conclusion can only be affected by entropy dilution arising from the decay 
of some heavy particles below the temperatures given in Eq.~(\ref{peak})~\cite{Asaka:06,Bezrukov:10}.

The production of sterile neutrino DM may substantially change in the presence of lepton asymmetry 
when the resonant production (\emph{RP}) of sterile neutrinos \cite{Shi:98} occurs, analogous to the 
Mikheyev--Smirnov--Wolfenstein effect~\cite{Wolfenstein:1977ue,Mikheev:1986gs}. When the dispersion 
relations for active and sterile neutrinos cross each other at some momentum $p$, the effective transfer 
of an excess of active neutrinos (or antineutrinos) to the population of DM sterile neutrinos occurs. 
The maximal amount of sterile neutrino DM that can be produced in such a way is 
limited by the value of 
lepton asymmetry, $\eta_L \equiv |n_{\nu}-n_{\bar\nu}|/s$, where $s$ is the entropy of relativistic species 
in plasma. The present DM abundance $\Omega_\dm \sim 0.25$ translates into the requirement of 
$\eta_L \sim 10^{-6}\Bigl(\frac{\mathrm{keV}}{M_{N_\dm}}\Bigr)$ in order for RP sterile neutrinos to 
constitute the dominant fraction of DM. One notices that the resonant production occurs only for values of 
lepton asymmetry, $\eta_L$ much larger than the measured value of baryon asymmetry of the Universe: 
$\eta_B \equiv \frac{n_B}{s} \sim 10^{-10}$~\cite{PlanckXVI:13}. Such a value of $\eta_L$ does not contradict to 
any observations though. Indeed, the upper bounds on $\eta_L$ are based on either primordial nucleosynthesis 
(BBN) or CMB measurements (as chemical potential of neutrinos would carry extra 
radiation density)~\cite{Lesgourgues:99,Kirilova:11}. These bounds read
$|\eta_L|\lesssim \text{few}\times 10^{-3}$ (see e.g.~\cite{Serpico:05,Mangano:10,Castorina:12}).
We see, therefore, that the lepton asymmetry, required for resonant sterile neutrino production is 
still considerably smaller than the upper limit. Notice, that at epochs prior to BBN even $\eta_L \sim 1$ 
is possible (if this lepton asymmetry disappears later). Such a scenario is realized e.g. in the 
\emph{Neutrino Minimal Standard Model}, $\nu$MSM (see~\cite{Boyarsky:09a} for review), 
where the lepton asymmetry keeps being generated below the sphaleron freeze-out temperature and may 
reach $\eta_L \sim 10^{-2} \div 10^{-1}$ before it disappears at $T\sim{}$~few GeV~\cite{Shaposhnikov:08a}.

\subsection{Structure formation with sterile neutrino dark matter}
\label{sec:structure-formation}

Non-negligible velocities of `warm' sterile neutrinos alter the power spectrum of density fluctuations at scales 
below the free-streaming horizon scale. Additionally, the suppression of the halo mass function below a certain 
scale~\cite{Benson:12} and different history of formation of first structures affects the way the first stars 
were formed and therefore the reionization history of the Universe, abundance of the oldest (\emph{Population III}) 
stars, etc~\cite{SommerLarsen:03,oShea:06,Gao:07b,Hansen:2003yj,Yoshida:03,Yue:2012na}.

The effects of suppression of the matter power spectrum are probed with the 
\textbf{Lyman-$\alpha$ forest method}~\cite{Hansen:01,Viel:05,Viel:06,Seljak:06,Viel:13} (see~\cite{Boyarsky:08c}
for critical overview of the method and up-to-date bounds). 
Using neutral hydrogen as a tracer of overall matter overdensity, one can reconstruct the power spectrum of 
density fluctuations at redshifts $2<z<5$ and scales $0.3-5$~h/Mpc (in comoving coordinates) 
by analyzing Lyman-$\alpha$ absorption features in the spectra of distant quasars.

If all DM is made of sterile neutrinos with a simple Fermi-Dirac-like spectrum of primordial 
velocities~(\ref{eq:19}), the matter power spectrum has a sharp (cut-off like) suppression 
(as compared to $\Lambda$CDM) at scales below the free-streaming horizon~(\ref{eq:5}) 
(similar to the case of `thermal relics'~\cite{Bode:00}). In this case the Lyman-$\alpha$ forest 
data~\cite{Hansen:01,Viel:05,Viel:06,Seljak:06,Viel:07,Boyarsky:08c} puts such strong constraints at their 
free-streaming length, which can be expressed as the \emph{lower bound} on their mass $M_{N_\dm} \ge 8\kev$
(at $3\sigma$ CL)~\cite{Boyarsky:08c}. Such WDM models produce essentially no observable changes in the 
Galactic structures (see~\cite{Strigari:06,Colin:07,Boyarsky:08c,deNaray:09,Schneider:11}) and therefore, 
from the observational point of view such a sterile neutrino DM (although formally `warm') 
would be indistinguishable from pure CDM.

On the other hand, resonantly produced sterile neutrinos have spectra that significantly differ 
from those in the non-resonant case~\cite{Shi:98,Laine:08a}. 
The primordial velocity distribution of RP sterile neutrinos contains narrow resonant (\emph{cold}) \emph{plus} 
a nonresonant (\emph{warm}) components -- CWDM model (see~\cite{Boyarsky:08c,Boyarsky:08d} for details).
\footnote{Axino and gravitino models may have similar spectra of primordial velocities, 
c.f.~\cite{Jedamzik:2005sx}.} In the CWDM case, however, Lyman-$\alpha$ constraints allow a significant 
fraction of DM particles to be very warm~\cite{Boyarsky:08c}. This result implies for example, 
that sterile neutrino with the mass as low as $1{-}2$~keV is consistent with all cosmological 
data~\cite{Boyarsky:08d}.

The first results~\cite{Lovell:11} demonstrate that RP sterile neutrino DM, compatible with the 
Lyman-$\alpha$ bounds~\cite{Boyarsky:08d}, do change the number of substructure of a 
Galaxy-size halo and their properties. Qualitatively, structures form in these models in a bottom-up fashion 
(similar to CDM). The way the scales are suppressed in CWDM models is more complicated 
(and in general less severe for the same masses of WDM particles), as comparable with pure warm DM models. 
The first results of~\cite{Lovell:11} demonstrate that the resonantly produced sterile neutrino DM models, 
compatible with the Lyman-$\alpha$ bounds of~\cite{Boyarsky:08d}, do change the number of substructure of a 
Galaxy-size halo and their properties. The discrepancy between the number of observed substructures 
with small masses and those predicted by $\Lambda$CDM models (first pointed out in~\cite{Klypin:99,Moore:99b}) 
can simply mean that these substructures did not confine gas and are therefore completely dark 
(see e.g.~\cite{Bullock:00,Benson:01b,Somerville:02,Maccio:10}). This is not true for larger objects. 
In particular, CDM numerical simulations invariably predict several satellites ``too big'' 
to be masked by galaxy formation processes, in contradiction with 
observations~\cite{Klypin:99,Moore:99b,Strigari:10,BoylanKolchin:11}. 
Resonantly produced sterile neutrino DM with its non-trivial velocity dispersion, 
turns out to be ``warm enough'' to amend these issues~\cite{Lovell:11} 
(and ``cold enough'' to be in agreement with Lyman-$\alpha$ bounds~\cite{Boyarsky:08d}).

Ultimate investigation of the influence of dark matter decays and of modifications in the evolution of 
large scale structure in the `sterile neutrino Universe' as compared with the $\Lambda$CDM model requires a 
holistic approach, where all aspects of the systems are examined within the same set-up rather than 
studying the influence of different features one-by-one. Potentially observable effects of particles' 
free streaming and decays are expected in terms of
\begin{compactitem}[--]
\item formation and nature of the first stars~\cite{Stasielak:06,Ripamonti:2006gr,Gao:07b,oShea:06};
\item reionization of the Universe~\cite{Hansen:2003yj,Biermann:06,Kusenko:06b,Mapelli:2006ej,Yue:2012na};
\item the structure of the intergalactic medium as probed by the Lyman-$\alpha$
  forest~\cite{Ripamonti:2006gq,Viel:12b,Viel:11,Viel:07,Viel:06,Seljak:06,Boyarsky:08c,Boyarsky:08d};
\item the structure of dark matter haloes as probed by gravitational 
lensing~\cite{Smith:11,Markovic:10,Viel:11,Miranda:07,Dunstan:11};
\item the structure and concentration of haloes of satellite 
galaxies~\cite{Lovell:11,Polisensky:10,Maccio:09,Maccio:12,Shao:12}.
\end{compactitem}

The results of this analysis will be confronted with measured cosmological observables, using various methods: 
Lyman-$\alpha$ analysis (with BOSS/SDSS-III~\cite{Schlegel:07} or 
X-Shooter/VLT~\cite{Xshooter}), statistics and structure 
of DM halos, gravitational lensing, cosmological surveys).

The weak lensing surveys can be used to probe further clustering properties of dark matter particles as 
sub-galactic scales, as the next generation of these surveys will be able to measure the matter power spectrum 
at scales down to $1-10$~h/Mpc with a few percent accuracy. The next generation of lensing surveys 
(such as e.g. KiDS~\cite{KiDS}, LSST~\cite{Abell:09}, WFIRST~\cite{Green:12}, 
Euclid~\cite{Beaulieu:10}) 
can provide sensitivity, compatible with the existing Lyman-$\alpha$ bounds~\cite{Markovic:10,Smith:11}. 
As in the case of the Lyman-$\alpha$ forest method the main challenge for the weak lensing is to properly 
take into account baryonic effects on matter power spectrum. The suppression of power spectrum due to 
primordial dark matter velocities can be extremely challenging to disentangle from the modification of 
the matter power spectrum due to baryonic feedback~\cite{Semboloni:11,vanDaalen:11,Viel:12b}. Finally, 
the modified concentration mass relation, predicted in the CWDM models, 
including those of resonantly produced sterile neutrinos (\cite{Boyarsky:08d,Maccio:12b}) 
can be probed with the weak lensing surveys (see e.g. \cite{Mandelbaum:08,King:11}) if their sensitivity 
can be pushed to halo masses below roughly $10^{12}M_\odot$.

\subsection{Sterile neutrinos as decaying dark matter}
\label{sec:decaying-dark-matter}

Sterile neutrino is an example of decaying dark matter candidate. 
The astrophysical search for decaying DM is very promising. 
First of all, a positive result would be conclusive, as the DM origin of any candidate signal 
can be unambiguously checked. Indeed, the decay signal is proportional to the \emph{column density} 
$S = \int\rho_\dm(r)dr$ along the line of sight and not to the $\int\rho^2_\dm(r)dr$ 
(as it is the case for annihilating DM). As a result, a vast variety of astrophysical objects of different 
nature would produce a comparable decay signal (c.f.~\cite{Boyarsky:06c,Boyarsky:09b,Boyarsky:09c}). 
Therefore 
\begin{inparaenum}[\em (i)]\item one has a freedom of choosing the observational targets, 
avoiding complicated astrophysical backgrounds; 
\item if e.g. a candidate spectral line is found, its surface brightness profile may be measured 
(as it does not decay quickly away from the centers of the objects), distinguished from astrophysical 
emissions (that usually decay in outskirts) and compared among several objects with the same expected signal. 
This allows to distinguish the decaying DM signal from any possible astrophysical background and 
therefore makes astrophysical search for the decaying DM another type of direct (rather than indirect) 
detection experiment.
\end{inparaenum}
The case of the astrophysical search for decaying DM has been presented in the recent 
White Papers~\cite{denHerder:09,Abazajian:09a}. This approach has been illustrated on the recent claim 
of~\cite{Loewenstein:09} that a spectral feature at $E\sim 2.5$~keV in the \textit{Chandra} observation 
of Willman~1 can be interpreted a DM decay line. 
Ref.~\cite{Boyarsky:10b} demonstrated that such an interpretation is ruled out by archival observations 
of M31 and Fornax/Sculptor dSphs with high significance (see also~\cite{Mirabal:10,Loewenstein:12}).
\footnote{We do not discuss here the
  claim~\cite{Prokhorov:10} that the intensity of the Fe XXVI Lyman-$\gamma$
  line at $8.7\kev$, observed in~\cite{Koyama:07} cannot be explained by
  standard ionization and recombination processes, and that the DM decay may
  be a possible explanation of this apparent excess. Spectral resolution of
  current missions does not allow to reach any conclusion. However, barring an
  \emph{exact} coincidence between energy of decay photon and Fe XXVI
  Lyman-$\gamma$, this claim may be tested with the new missions, discussed
  in~\ref{sec:x-ray-calorimeters}.}

The `Tremaine–Gunn bound' restricts the lowest energies in which one can search for the fermionic decaying DM 
to the \emph{X-ray range}. An extensive search of the DM decay signal in the keV range using archive data 
was conducted recently, using \emph{XMM-Newton}, \emph{Chandra} and \textit{Suzaku} observations of 
extragalactic diffuse X-ray background, galaxies and galaxy 
clusters~\cite{Boyarsky:05,Riemer:06,Boyarsky:06e,Abazajian:06,Abazajian:06b,Riemer:09a,Boyarsky:06b,Boyarsky:06c,Boyarsky:06d,Watson:06,Boyarsky:07a,Boyarsky:07b,Loewenstein:08}. 
This search allowed to probe large part of the parameter space of decaying DM (between 0.5~keV and $\sim$14~MeV) 
and establish a \emph{lower bound} on the lifetime of dark matter decay for both $\text{DM} \to \nu + \gamma$ and
also $\text{DM} \to \gamma + \gamma$ (the latter would be the case e.g. for axion or majoron~\cite{Lattanzi:07}). 
The combined restrictions on the lifetime (see~\cite{Boyarsky:08b}) turns out to exceed $10^{26}$~s, 
almost independent on the mass.

\begin{figure}[t]
  \centering
  \includegraphics[width=0.75\textwidth]{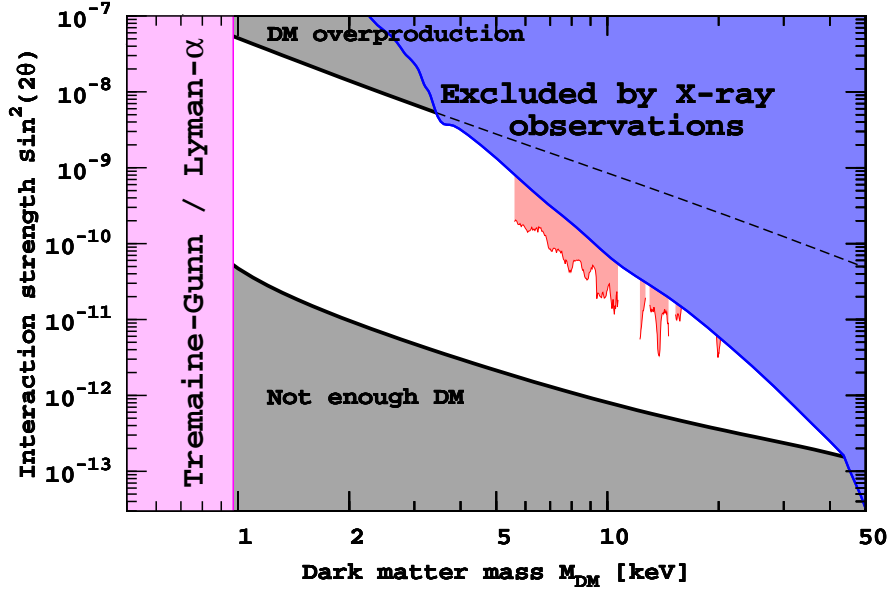}
  \caption{The allowed region of parameters of sterile neutrino dark matter in
    the \numsm (white unshaded region) confronted with existing and projected
    experimental bounds.  For any combination of mass and mixing angle between
    two black curves the necessary amount of dark matter can be produced
    (given the presence of certain amount of lepton asymmetry in the plasma,
    generated by two other sterile neutrinos).  The blue shaded region in the
    upper right corner is excluded by the non-observation of decaying DM line
    in
    X-rays~\cite{Boyarsky:05,Boyarsky:06c,Boyarsky:06d,Watson:06,Boyarsky:07a,Abazajian:06,Riemer:06,Boyarsky:07b,Loewenstein:08}.
    Red regions between $\sim 5\kev$ and $\sim 20\kev$ show \emph{expected
      sensitivity} from a \emph{combination of a large number of archival
      observations} (as described in Section~\ref{sec:stack-observ}). The gaps
    are due to the presence of strong instrumental lines at certain energies
    (where the combination method does not provide any improvement over
    earlier bounds).  The lower limit of $\sim 5\kev$ is due to the presence
    of instrumental lines and absorption edge at energies $1-2.5\kev$ and
    emission of the Milky way, dominating at lower energies.  In the region
    below $1\kev$ sterile neutrino DM is `too light' and is ruled out based on
    'Tremaine-Gunn' like arguments~\cite{Boyarsky:08a} and on the
    Lyman-$\alpha$ analysis~\cite{Boyarsky:08c,Boyarsky:08d}.}
  \label{fig1}
\end{figure}

Let us consider the implications of the negative results of searches for decaying dark matter line in the \numsm, 
taking it as a minimal (baseline) model. Its parameter space is presented in Fig.~\ref{fig1}. 
For any combination of mass and mixing angle between two black curves the necessary amount of dark matter 
can be produced (given the presence of certain amount of lepton asymmetry in the plasma). 
If interaction strength is too high, too much dark matter is produced in contradiction with observations. 
If the interaction strength is too low -- one cannot account for 100\% of dark matter with sterile neutrinos 
and additional ``dark'' particles would be needed). The shaded region in the upper right corner is excluded 
due to non-observation of decaying dark matter line with X-ray 
observatories~\cite{Boyarsky:05,Riemer:06,Boyarsky:06e,Abazajian:06,Abazajian:06b,Riemer:09a,Boyarsky:06b,Boyarsky:06c,Boyarsky:06d,Watson:06,Boyarsky:07a,Boyarsky:07b,Loewenstein:08}. 
Confronting the requirement to produce the correct DM abundance with the X-ray bounds, 
one is able to deduce the upper limit on the mass of sterile neutrino DM to be about 50~keV~\cite{Boyarsky:07b}. 
Finally, a lower limit on the mass of DM sterile neutrino $M_{N} \sim 1-2$~keV comes from the analysis of the 
Lyman-$\alpha$ forest data~\cite{Boyarsky:08c,Boyarsky:08d}.\footnote{Notice, that the lower bound
  on the mass of sterile neutrino DM, produced via non-resonant mixing
  (having a simple Fermi-Dirac-like spectrum) is at tension with the upper
  bound on the mass, coming from X-ray observations (see
  e.g.~\cite{Boyarsky:07a,Boyarsky:08c} and refs.\ therein).} 
As a result, the combination of X-ray bounds and computations of primordial abundance shows that in the 
\numsm the parameter space of sterile neutrino DM is \emph{bounded on all sides}.

To further advance into the allowed region of the \numsm  (the simplest model, predicting sterile neutrino DM) 
one has either drastically improve the statistics of observations of DM-dominated objects 
(Section~\ref{sec:stack-observ}), or employ new technologies of detecting X-rays in space that deliver better 
spectral resolution than existing X-ray missions (Section~\ref{sec:x-ray-calorimeters}).

\subsection{Advance with existing missions: stacking of observations}
\label{sec:stack-observ}

Significant improvement of sensitivity for decaying DM line with the current X-ray missions 
(\textit{XMM-Newton, Chandra, Suzaku}) is quite challenging. Indeed, an improvement by an order of 
magnitude would require an increase of observational time by \emph{two} orders of magnitude. 
The best existing constraints in X-rays are based on observations with exposure of several hundreds of ks. 
Therefore, one would need $\gtrsim 10$~Ms of dedicated X-ray observations. 
Such a huge cleaned exposure is extremely difficult to obtain for a \emph{single} DM-dominated object 
(for example, the whole year of observational programme of the \textit{XMM-Newton} satellite is only 14.5~Ms).

Using archive of the \textit{XMM-Newton} observations\footnote{\textit{XMM-Newton} has the largest `grasp' 
(=product of the field-of-view and effective area) as compared to \textit{Chandra} and \textit{Suzaku}, 
which would allow to collect the largest amount of photons from `diffuse sources', such as the signal 
from DM decays in the DM halos of the nearby galaxies.} it is possible to collect about 20~Ms of 
observations of nearby spiral and irregular galaxies~\cite{Boyarsky:12c} (galaxy clusters have much stronger 
emission in the keV range and their combined analysis would require a completely different strategy). 
Therefore a possible way to advance with the existing X-ray instruments is to combine a large number 
of X-ray observations of different DM-dominated objects. The idea is that the spectral position of the DM 
decay line is the same for all these observations, while the astrophysical backgrounds in the combined spectrum 
would ``average out'', producing a smooth continuum against which a small line would become visible. 
Naively, this should allow to improve the existing bounds by at least an order of magnitude.

\begin{figure}[t]
  \centering
  \includegraphics[width=0.75\linewidth]{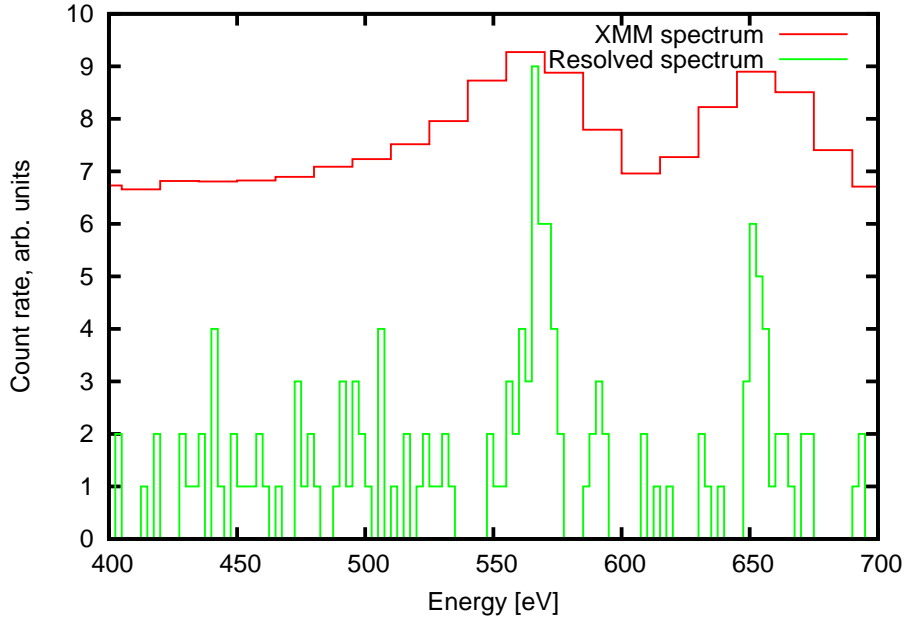}
  \caption{Galactic diffuse background (observed with \textit{XMM-Newton}
    (red) and the same data, observed with the X-ray spectrometer (XQC
    project~\cite{McCammon:02}).}
  \label{fig:resolved-spectra}
\end{figure}

However, this turns out to be a highly non-trivial task. Indeed, such a large exposure means that the statistical 
errors in each energy bin can be as small as 0.1\%. To extract meaningful bounds one would need therefore 
\emph{comparably small} systematic errors. However, the level of systematics of the \textit{XMM-Newton} 
is much higher (at the level 5--10\%, see e.g.~\cite{XMM-SOC-TN-0018}) due to the instrument's degradation 
with time and variability of the instrumental (=cosmic-ray induced) background that constitutes a significant 
part of a signal in each energy bin (and becomes a dominant component above few keV 
(c.f.~\cite{Read:03,Nevalainen:05,Carter:07,Kuntz:08}). The exposure of `closed filter' 
dataset\footnote{A special dataset (obtained with the filter of the X-ray telescope closed, so that no 
X-ray photons can reach the detector) created specifically to determine the (time-averaged) shape of the 
instrumental background and used to remove the most prominent instrumental features from observations of 
diffuse sources, see e.g.~\cite{Kuntz:08,Lumb:02}.} is $\sim 1$~Ms. As a result, 
the usual practice of subtraction of rescaled instrumental background data 
(see e.g.~\cite{Kuntz:08,Carter:07,Pradas:05}) would mean at least $\sim 3$ times larger errorbars due to the 
smaller exposure of the instrumental dataset. Moreover, the instrumental component of the \textit{XMM-Newton} 
background is self-similar only on average which would introduce additional errors 
(at the level of few \%, see~\cite{Boyarsky:06d}). Another standard procedure of working with diffuse 
sources -- subtraction of the `blank sky' data\footnote{Combination of many observations of X-ray quiet parts 
of the sky~\cite{Carter:07,Read:03}. Unlike the `closed filter' dataset collects the physical emission from the 
Milky Way.} will not be applicable in this case as well. First of all, such a dataset would also contain decaying 
dark matter line originating from the decays in the Milky Way halo 
(this fact has been exploited before to put bounds on decaying DM in~\cite{Boyarsky:06d,Abazajian:06}). 
Secondly, subtracting the `blank sky' data would again reduce all the advantages of a large dataset by 
lowering statistics (as the exposure of the latest blank-sky co-added observations is again of the order 
of $\sim 1$~Ms).

This means that to \emph{take all the advantages of this long-exposure dataset, one cannot use 
the standard data-processing methods}. Therefore, an alternative method of data analysis has to be developed, 
that has the sensitivity towards the searching for narrow lines at the level, dictated by the statistics 
of the combined dataset. The results will be reported in~\cite{Boyarsky:12c}. The estimated level of sensitivity 
of this method is shown as the red line in Fig.~\ref{fig1}.

\begin{figure*}[t]
  \centering
  \includegraphics[width=.5\textwidth]{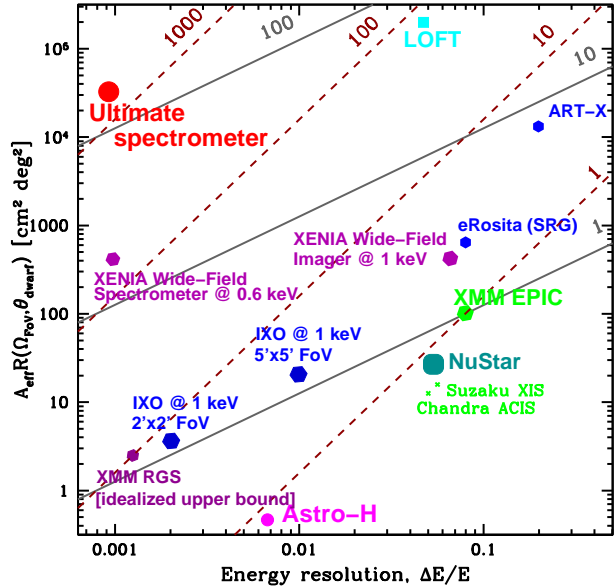}
  \caption{Comparison of sensitivities of existing and proposed/planned X-ray
    missions for the detection of the DM decay line in a nearby dwarf
    spheroidal galaxy of the angular size of $1\deg$.  The sensitivity of
    XMM-Newton EPIC camera is taken as a reference.  Solid lines indicate
    improvement of the sensitivity by factors of 1, 10 and 100 (the top left
    is the most sensitive).  The dashed lines show the improvement of the
    sensitivity towards the detection of a strong line (in an effectively
    background free regime). See also~\cite{Boyarsky:06f,denHerder:09}. }
   \label{fig:missions}
\end{figure*}

\subsection{X-ray micro-calorimeters}
\label{sec:x-ray-calorimeters}

Really significant progress (that allows, for example, to cover the whole region of parameter space 
in Fig.~\ref{fig1}) in searching for decaying DM cannot be achieved with the existing instruments by 
simply increasing the exposure of observations. Indeed, the width of the DM decay line, ${\Delta}E/E_{\gamma}$ 
is determined by the virial velocities of DM particles in halos and ranges from $\mathcal{O}(10^{-4})$ for dwarf 
spheroidal galaxies to $\mathcal{O}(10^{-3})$ for the Milky Way-size galaxies to $10^{-2}$ for galaxy clusters. 
If the spectral resolution is much bigger than the width of the line, one averages the photons 
from the line with the background photons over a large energy bin. This is the case for all existing 
X-ray missions, whose detectors are based on CCD technology (c.f.~\cite{CCD}) and where the spectral 
resolution is at the level $\Delta E/E \gtrsim 10^{-2}$, see Fig.~\ref{fig:missions}. Therefore, 
\emph{an X-ray spectrometer with the energy resolution at least $\Delta E/E \sim 10^{-3}$ is crucial for 
detection of a decaying DM line}.

%%%%%%%%%%%%%%%%%%%%%%%%%%%%%%%%%%%%%%%%%%%%%%%%

The technology behind such spectrometers (known as \emph{X-ray micro-calorimeters}, 
see e.g.~\cite{Porter:04,McCammon:05}) has been actively developed by the high-energy astrophysical community 
in the last decades. There is a strong interest for building such a spectrometer, 
and different versions of high resolution X-ray missions had been proposed in response to the ESA and NASA calls 
(including the ESA's call for Fundamental Physics Roadmap), see 
e.g.~\cite{denHerder:09,Piro:08,Astro-H,Herder:11,Athena}. Astrophysical interest to X-ray spectrometer 
is motivated by a number of important applications to observational cosmology, 
providing crucial insight into the nature of dark matter by studying the structure of the ``cosmic web''. 
In particular, 
\begin{inparaenum}[(i)]
\item search for missing baryons in the cosmic filaments; through their emission and absorption; 
\item trace the evolution and physics of clusters out to their formation epoch; 
\item gamma-ray bursts as a source of backlight to observe the warm-hot intergalactic media in absorption; 
\item study the evolution of massive star formation using gamma-ray bursts to trace their explosions back 
to the early epochs of the Universe ($z \sim 6$) (see e.g.~\cite{Piro:08,Herder:11,Athena}). 
\end{inparaenum}

The first spectrometer based on this technology was flown (albeit unsuccessfully) on \textit{Suzaku} 
mission~\cite{Kelley:06} and another one is being planned for the Astro-H~\cite{AstroH,Astro-H} 
(to be launched in 2014). However, currently planned and proposed X-ray micro-calorimeter missions 
(Astro-H~\cite{Astro-H}, Athena~\cite{Athena}, ORIGIN~\cite{Herder:11}, etc.) are not optimal for the 
purpose of decaying dark matter search. These missions are optimized for the astrophysical goals and have 
limited field-of-view (usually, much below $\unit[1]{deg}^2$), good angular resolution and narrow energy range.

On the contrary, the key parameters that determine the sensitivity of the proposed instrument for decaying 
dark matter search are (see Fig.~\ref{fig:missions}):
\begin{compactitem}[--]
\item a spectral resolution $\Delta E/E \lesssim 10^{-3}$ over the range of energies $\unit[0.5 - 25]{keV}$ 
(this is the minimal energy range, that would allow to probe the parameter space of our baseline model, the \numsm);
\item large `\emph{grasp}' $\sim \unit[10^3-10^4]{cm^2\times deg^2}$. There are essentially two possibilities 
to achieve such a grasp. One can either launch a non-imaging spectrometer 
(with a `collimator' having a field-of-view as large as $\sim \unit[10^2]{deg}^2$)\footnote{Making field-of-view 
significantly larger than about $10^\circ \times 10^\circ$ would of course further increase the sensitivity 
towards the line detection. However, in this case it would become challenging to identify the nature of the 
candidate line (if found), as in this case none of the nearby DM dominated objects with large angular size 
(Andromeda galaxy, Large and Small Magellanic clouds, Virgo cluster) will look like `hot spots' of DM decays. 
Moreover, in this case it will not be possible to build a DM surface brightness profile as one varies the 
directions off the Galactic Center and investigate whether it is consistent with DM distribution in the Milky 
Way. }; or install mirrors (thus increasing the effective area beyond the geometric size of the detectors, 
probably to as much as $\unit[10^3]{cm^2}$). The latter option allows to have also imaging capabilities, however, 
it is usually extremely costly to cover the required energy range and to have sufficiently large 
(at least $1^\circ \times 1^\circ$) field of view.
  \end{compactitem}
Fig.~\ref{fig:missions} summarizes sensitivity of existing and proposed missions and demonstrates that none 
of them would provide a sufficient improvement with respect to the existing constraints 
(see~\cite{Boyarsky:06f,denHerder:09} for discussion).

Currently, there exists a project (the \emph{X-ray quantum calorimeter}, XQC~\cite{McCammon:02}) that can be 
considered a prototype of the proposed mission. It has the field of view of about $\unit[1]{sr}$ 
($\unit[3.5\times 10^3]{deg^2}$), an effective area of $\sim \unit[1]{cm^2}$ and the energy resolution of 
10~eV over the energy range 0.1-4~keV~\cite{McCammon:02}.\footnote{A similar calorimeter used in Suzaku was
  capable of delivering a similar resolution up to the maximal energy range of  $12\kev$~\cite{Kelley:06}.} 
This calorimeter has been flown several times on sounding rockets~\cite{McCammon:02}. Although each flight 
had been very short (about 100~s), it allowed to demonstrate that the Milky Way emission in the energy 
range $\unit[0.1 - 1]{keV}$ (which looks as a continuum in the spectra obtained with X-ray imaging instruments, 
see e.g.~\cite{Markevitch:03,Carter:07} is actually a ``forest'' of thin lines 
(see Fig.~\ref{fig:resolved-spectra}). Because of its superior spectral resolution, 
decaying DM bounds based on the $\sim 100$~s exposure of the flight of this 
spectrometer~\cite{McCammon:02} are comparable with $10^4$~s of the \textit{XMM-Newton} 
exposure~\cite{Boyarsky:06f}.

To detect a dark matter decay line, that is much weaker than the lines resolved with the XQC spectrometer, 
a significantly longer exposure ($\sim\unit[1]{year}$) would be required. The requirement to keep the 
cryostat of such a spectrometer in the stable regime, means that one cannot use the sounding rockets, 
but rather needs to use a satellite (probably, staying in Low Earth Orbit, unlike \textit{XMM-Newton} or 
\textit{Chandra}). The project therefore becomes a small-to-medium scale cosmic mission.

\subsection{Laboratory searches for sterile neutrino DM}
\label{sec:non-astr-search}

Finally, several words should be said about laboratory searches of DM sterile neutrinos. 
As Fig.~\ref{fig1} demonstrates, their mixing angle is always smaller than $\sim 10^{-4}$
(even for the lightest admissible masses of $\sim 1\kev$). This makes their laboratory searches 
extremely challenging. One possibility would be to measure the event-by-event kinematics of 
$\beta$-decay products~\cite{Bezrukov:06}. This experiment, however, is plagued by the bremsstrahlung 
emission of the finite state electrons that changes their energy. Other possibilities of searches for 
the keV-scale sterile neutrinos are discussed e.g. in~\cite{Ando:10,Liao:10,deVega:2011xh}. 
All these experiments require essentially background-free regime and it is not clear whether any 
of them can realistically touch cosmologically interesting region of parameters of sterile neutrino.

\section{Accelerator searches for sterile neutrinos: present and future}
\label{sec:searching-sterile-neutrinos}

Although only one sterile neutrino plays the role of dark matter, the fact that three of them are needed 
to explain \emph{both} dark matter and neutrino oscillations is crucial as these two particles set up 
\emph{the initial conditions} for sterile neutrino DM production and affect their primordial 
properties~\cite{Shi:98,Laine:08a,Shaposhnikov:08a,Canetti:12}. If the masses of sterile neutrinos 
responsible for neutrino oscillations are below electroweak scale (as it is the case in the \numsm), 
such particles can be found in `intensity frontier' experiments, opening the road for the experimental 
resolution of three major observational problems `beyond-the-Standard-Model': neutrino flavor oscillations, 
matter-antimatter asymmetry of the Universe and dark matter.

\subsection{Direct searches for sterile neutrinos with MeV--GeV masses}
\label{sec:accelerator-searches}

The idea that the SM can be extended in the neutrino sector by adding several relatively light neutral 
fermions was discussed intensively since the 1980s. Two distinct strategies have been used for these searches. 
The first one is related to their production. The neutral leptons participate in all reactions the ordinary 
neutrinos do with a probability suppressed by their mixing angles with active neutrinos. 
Since sterile neutrinos are massive, 
the kinematics of 2-body decays $K^\pm \rightarrow \mu^\pm N$, $K^\pm \rightarrow e^\pm N$ 
or 3-body decays $K_{L,S}\rightarrow \pi^\pm + e^\mp + N$ changes when $N$ is replaced by an ordinary 
neutrino~\cite{Shrock:80}. Therefore, the study of \emph{kinematics} of rare meson decays can constrain the 
strength of the coupling of heavy leptons. This strategy has been used for the search of neutral leptons in 
the past, where the spectrum of electrons or muons originating in decays $\pi$ and $K$ mesons has been 
studied~(\cite{Britton:1992pg,Britton:1992xv,Yamazaki:1984sj,Bryman:1996xd,Abela:1981nf,Daum:1987bg,Hayano:1982wu}
see discussion in \cite{Asaka:11a,Ruchayskiy:11a,Lello:12}). The second strategy is to look for the decays 
of sterile neutrinos\footnote{These sterile
  neutrinos are also produced in the weak decays of heavy mesons and
  baryons. From the experimental point of view, the distinct mass ranges are
  associated with the masses of parent mesons: below $\unit[500]{MeV}$ (K
  meson), between $\unit[500]{MeV}$ and $2$ GeV (D-mesons), between $2$ and
  $5$ GeV (B-mesons), and above $5$ GeV. (see~\cite{Gorbunov:07a,gorbunov} for
  details).} % 
inside a detector (`\emph{nothing}' $\rightarrow$ leptons and hadrons). Typical patterns and branching ratios 
of sterile neutrino decays can be found in~\cite{Gorbunov:07a}.

Such searches have beeen undertaken at CERN, FNAL, PSI and other laboratories 
(e.g.\ PS191~\cite{Bernardi:1985ny,Bernardi:1987ek,Vannucci:2008zz}, BEBC, CHARM\cite{CHARM:1985}, 
NOMAD\cite{Astier:2001ck} and NuTeV~\cite{Vaitaitis:1999wq},
see~\cite{Kusenko:2004qc,Gorbunov:07a,Atre:09,Ruchayskiy:11a} for review).
However only recently understanding that the \emph{light singlet fermions in the region of accessibility of 
existing accelerators can explain neutrino oscillations allowed to fix an ultimate goal for searches of 
neutral leptons}.

%%%%%%%%%%%%%%%%%%%%%%%%%%%%%%%%%%%%%%%%%%%%%%%%%
\begin{figure}[t]
  \centering
  \includegraphics[width=0.47\textwidth]{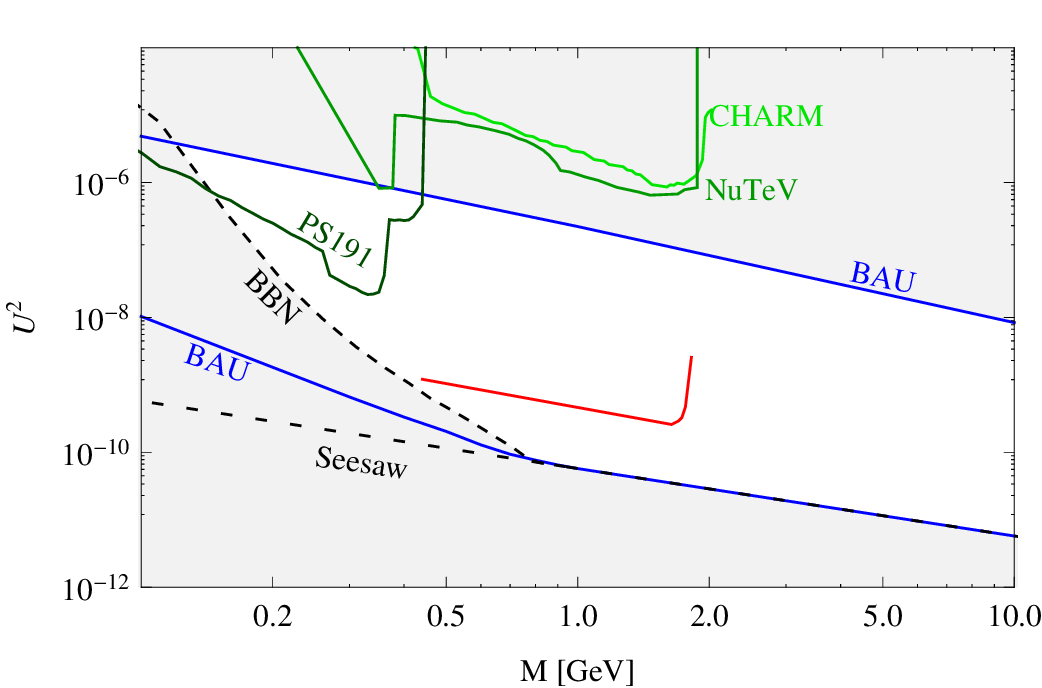}~~
  \includegraphics[width=0.47\textwidth]{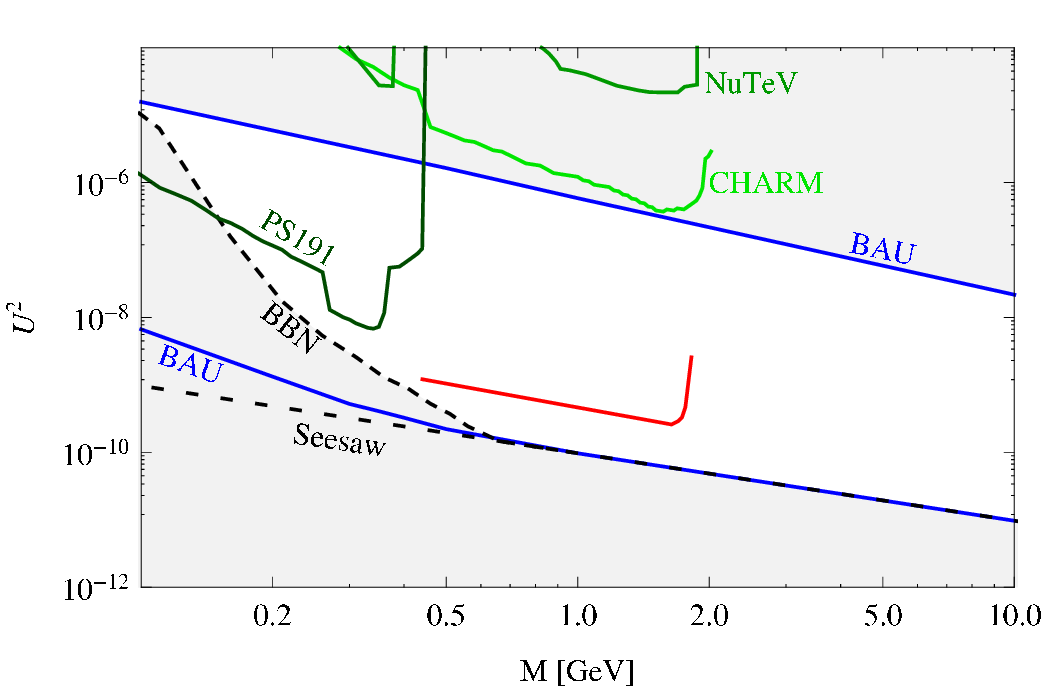}
  \caption{The allowed region of parameters of sterile neutrinos (mass vs.\
    sum of the mixing angles, $U^2 = \theta^2_2+\theta^2_3$). Parameters of
    neutrinos, responsible for neutrino oscillations are in the region above
    the dotted ``see-saw'' line.  Successful baryo/leptogenesis is possible in
    the region between two black solid lines. Different points within the
    allowed region correspond to different choices of a (unknown) CP-violating
    phase of active-sterile Yukawa matrix.  Sterile neutrinos with the
    parameters in the shaded region to the left of the ``BBN'' line would
    spoil predictions of primordial nucleosynthesis (based
    on~\cite{Dolgov:00b,Dolgov:00a}).  Accelerator
    experiments~\cite{Bernardi:1985ny,Bernardi:1987ek,Vannucci:2008zz,CHARM:1985,Astier:2001ck,Vaitaitis:1999wq},
    searching for heavy neutral leptons exclude regions above green
    lines. {\it Left panel:} restrictions for normal hierarchy, {\it Right
      panel:} inverted hierarchy. Adopted from Ref. \cite{Canetti:12}. The
    region above the red curve can be probed {\bf with a single section of the
      detector} similar to the one used in PS191 experiment but of large size
    (length $l_{||}\sim \unit[100]{m}$, height $\unit[5]m$ and width
    $l_\bot\sim\unit[5]m$, placed at a distance of about hundred meters from a
    beam target; see the proposal~\cite{gorbunov} for details. Adopted
    from~\cite{Canetti:12}.}
\label{exp}
\end{figure}
%%%%%%%%%%%%%%%%%%%%%%%%%%%%%%%%%%%%%%%%%%%%%%%%%%

Moreover, sterile neutrinos with such parameters can also provide an explanation of \emph{the 
observed matter-antimatter asymmetry} of the Universe~\cite{Asaka:05a,Canetti:10a,Canetti:12}, and therefore 
if these particles are found, one receives \emph{a unique possibility of direct experimental verification of 
the mechanism of baryogenesis}, checking if the parameters of the found sterile neutrinos satisfy the 
requirements of successful baryogenesis. Out of experiments already made, only 
CERN PS191~\cite{Vannucci:2008zz} entered deeply into the cosmologically interesting part of parameters 
for the masses of singlet fermions below that of kaon.

The lower limit on the mass of these particles is determined by combination of particle physics and 
cosmological considerations and should be above $\sim 100$ MeV~\cite{Ruchayskiy:11a,Ruchayskiy:12a} 
while no known solid upper bound, better than the electroweak scale, can be applied. At the same time, 
various considerations indicate that their mass may be in ${\cal O}(1)$ GeV 
region~\cite{Shaposhnikov:08a,Boyarsky:12b}.

\subsection{Future searches for sterile neutrinos with intensity frontier
  experiments}
\label{sec:method}

A future ``intensity frontier'' experiments (such as NA62~\cite{NA62}, measuring very rare kaon 
decay $K\to \pi \nu \bar\nu$; Long-Baseline Neutrino Experiment (LBNE)~\cite{LBNE,LBNE:11}; etc.) 
or even modifications of some of the existing experiments (such as e.g. T2K, see~\cite{gorbunov}) 
would be able to enter cosmologically interesting region of the sterile neutrino parameter space shown 
in Fig.~\ref{exp}. An experimental setup can be the following. Heavy mesons (and baryons) are produced in 
``fixed target experiments'' (beam of energetic protons hitting a target). 
Sterile neutrinos are created in the decays of these mesons and one can then search for their decays 
into pairs of charged particles (the probability of production and subsequent decay is 
proportional to $\theta_\alpha^2 \times \theta_\beta^2$, where $\alpha$ and $\beta$ are flavor indexes).

\begin{figure}[!t]
  \hskip 0.05\textwidth
\includegraphics[width=0.9\textwidth]{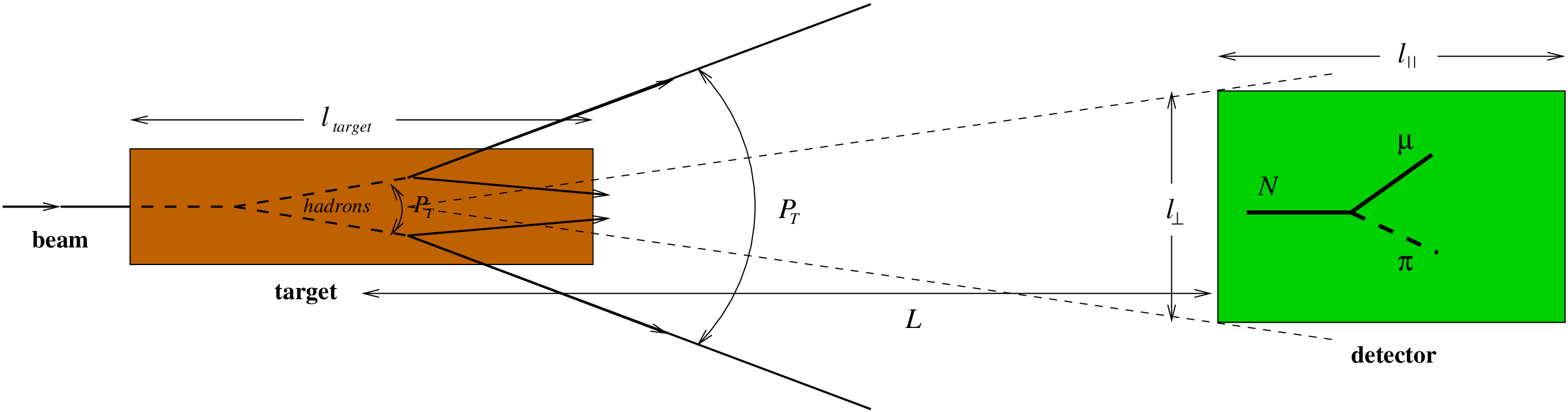}
\caption{Sketch of typical beam-target experiment on searches for sterile
  neutrino decays. Heavy mesons (and baryons) can be produced by energetic
  protons scattering off the target material.  Sterile-active neutrino mixing
  gives rise to sterile neutrino production in weak decays of the heavy
  mesons. These sterile neutrinos further weakly decay to the SM particles due
  to mixing with active neutrinos. Before the decay (and with account of
  $\gamma$-factor) relativistic sterile neutrinos would cover quite a large
  distance significantly exceeding ten kilometers. From~\cite{gorbunov}.}
\label{setup}
\end{figure}

Before the decay (taking into account their Lorentz factor) sterile neutrinos would travel large distances 
($c\tau_N \sim \unit[\mathcal{O}(10)]{km}$). Hence, sterile neutrino decays into SM particles due to 
mixing with active neutrino can be searched for in the near detector, see Fig.~\ref{setup}.

It is feasible to \emph{fully explore} the $\nu$MSM parameter space for sterile neutrino masses in the 
interesting range $M_N\sim0.5-2$\,GeV, where sterile neutrinos are dominantly produced in charmed hadron decays, 
one needs the following configuration (see~\cite{gorbunov,Gninenko:13} for details): 
\begin{compactitem}[--]%[\it (a)]
\item high intensity proton beam with about $10^{20}$ protons incident on the target per year; 
\item near detector, having the size $5$\,m$\times5$\,m$\times100$\,m and placed at a distance of about 
$1~\unit{km}$.\footnote{To register all the expected sterile neutrino decays
    one may consider installing many detectors of a reasonable size rather
    than one large detector (see~\cite{gorbunov} for details).}
\item Each detector may be relatively cheap, empty-space with simple tracker system inside and calorimeter 
at the far end. Its design may repeat the design of the CHARM experiment on searches for sterile 
neutrino decays at CERN SPS beam~\cite{CHARM:1985}. 
\end{compactitem}

At the mass range above 2~GeV the searches become more difficult, as the intensity of proposed flavor 
factories does not seem to be enough to collect sufficient amount of hadrons~\cite{Gorbunov:07a,gorbunov} 
(see e.g.~\cite{SuperB} for an overview of intensities of planned flavor factories). 
Some part of the parameter space of sterile neutrinos below $b$-quark mass (5~GeV) can be probed with the 
upgrade of LHCb experiment, see~\cite{LHCb-upgrade}, especially Section~2.2.1 there.

Additionally, experiments, searching for $\mu\to e\gamma$ and $\mu \to eee$ decays may touch the cosmologically 
interesting region of the parameter space of sterile neutrinos at masses above 10~GeV~\cite{Alonso:12}.

Finally, we notice, that the neutrinoless double-beta decay ($0\nu\beta\beta$) does not provide significant 
restrictions on the parameters of these sterile neutrinos (contrary to the case discussed in e.g.~\cite{Atre:09}), 
see discussion in~\cite{Blennow:2010th,Asaka:11a,Ruchayskiy:11a}. In particular, this is the case in the 
\numsm~\cite{Bezrukov:2005mx}, where the mass limits are 
$1.3~{\rm meV} <m_{\beta\beta}< 3.4~{\rm meV}$ ($13~{\rm meV}< m_{\beta\beta}< 50~{\rm meV}$) for normal 
(inverted) hierarchy. Detection of $m_{\beta\beta}$ outside these ranges would rule out the simplest model 
with only two sterile neutrinos with the masses in MeV--GeV range, responsible for neutrino oscillations.

\section{Conclusion}
\label{sec:conclusion}

After almost 20 years of research sterile neutrino remains a viable dark matter candidate. 
Observations of neutrino flavor oscillations further increased the interest to this candidate. 
Recent discovery of a Higgs like particle with the mass 125--126~GeV and absence of signs of new physics at 
the LHC or in DM direct detection experiments call for alternative (not related to electroweak symmetry breaking) 
testable beyond the Standard Model (BSM) models (including dark matter). Attempts to solve all BSM problems with particles with masses 
below electroweak scale~\cite{Shaposhnikov:07b,Boyarsky:09a} provide an novel approach to the problem of 
naturalness of the SM.

Dedicated cosmic experiment, an X-ray spectrometer, searching for signatures of decaying dark matter, 
has a capability to identify the dark matter particle. Combination of this experiment with the searches for 
neutral leptons at beam-target experiments gives a unique possibility to resolve experimentally three major 
BSM problems: the nature of neutrino flavor oscillations; the mechanism of generation of matter-antimatter 
asymmetry in the Universe; and the existence of dark matter. It could provide not only a possibility to 
detect new particles, but also do independent cross checks of the mechanisms of DM production and baryogenesis. 
Even negative results would allow to shed a light on the DM properties and therefore restrict the class of 
extensions of the SM.

Although current data, describing formation of structures, is fully consistent with the $\Lambda$CDM 
`concordance' model, sterile neutrino DM (that can be `warm', `cold' or `mixed' (cold+warm)) is also fully 
compatible with the observations. Future cosmic surveys will be able to measure the matter power spectrum 
with the sufficiently high precision to detect the imprints that such DM leaves in the matter power 
spectrum at sub-Mpc scales.

\subsubsection*{Acknowledgments}

This work was supported in part by the Swiss National Science Foundation. D.~I. also acknowledges support 
from the ERC Advanced Grant 20080109304, SCOPES Project IZ73Z0\_128040 of Swiss National Science Foundation, 
Grant No.~CM-203-2012 for young scientists of National Academy of Sciences of Ukraine, Grant 
No.~GP/F44/088 for talented young scientists of President of Ukraine, 
Cosmomicrophysics programme of the National Academy of Sciences of Ukraine and 
State Programme of Implementation of Grid Technology in Ukraine.

\singlespace

% \bibliographystyle{JHEP-2}
% \bibliographystyle{naturemag}
% 
% \bibliography{preamble,astro,cmf,combined_numsm,artem,strategy,combination_bib,calorimeter_review,thesis_Dima-reduced} %

\let\jnlstyle=\rm\def\jref#1{{\jnlstyle#1}}\def\aj{\jref{AJ}}
  \def\araa{\jref{ARA\&A}} \def\apj{\jref{ApJ}\ } \def\apjl{\jref{ApJ}\ }
  \def\apjs{\jref{ApJS}} \def\ao{\jref{Appl.~Opt.}} \def\apss{\jref{Ap\&SS}}
  \def\aap{\jref{A\&A}} \def\aapr{\jref{A\&A~Rev.}} \def\aaps{\jref{A\&AS}}
  \def\azh{\jref{AZh}} \def\baas{\jref{BAAS}} \def\jrasc{\jref{JRASC}}
  \def\memras{\jref{MmRAS}} \def\mnras{\jref{MNRAS}\ }
  \def\pra{\jref{Phys.~Rev.~A}\ } \def\prb{\jref{Phys.~Rev.~B}\ }
  \def\prc{\jref{Phys.~Rev.~C}\ } \def\prd{\jref{Phys.~Rev.~D}\ }
  \def\pre{\jref{Phys.~Rev.~E}} \def\prl{\jref{Phys.~Rev.~Lett.}}
  \def\pasp{\jref{PASP}} \def\pasj{\jref{PASJ}} \def\qjras{\jref{QJRAS}}
  \def\skytel{\jref{S\&T}} \def\solphys{\jref{Sol.~Phys.}}
  \def\sovast{\jref{Soviet~Ast.}} \def\ssr{\jref{Space~Sci.~Rev.}}
  \def\zap{\jref{ZAp}} \def\nat{\jref{Nature}\ } \def\iaucirc{\jref{IAU~Circ.}}
  \def\aplett{\jref{Astrophys.~Lett.}}
  \def\apspr{\jref{Astrophys.~Space~Phys.~Res.}}
  \def\bain{\jref{Bull.~Astron.~Inst.~Netherlands}}
  \def\fcp{\jref{Fund.~Cosmic~Phys.}} \def\gca{\jref{Geochim.~Cosmochim.~Acta}}
  \def\grl{\jref{Geophys.~Res.~Lett.}} \def\jcp{\jref{J.~Chem.~Phys.}}
  \def\jgr{\jref{J.~Geophys.~Res.}}
  \def\jqsrt{\jref{J.~Quant.~Spec.~Radiat.~Transf.}}
  \def\memsai{\jref{Mem.~Soc.~Astron.~Italiana}}
  \def\nphysa{\jref{Nucl.~Phys.~A}} \def\physrep{\jref{Phys.~Rep.}}
  \def\physscr{\jref{Phys.~Scr}} \def\planss{\jref{Planet.~Space~Sci.}}
  \def\procspie{\jref{Proc.~SPIE}} \let\astap=\aap \let\apjlett=\apjl
  \let\apjsupp=\apjs \let\applopt=\ao \def\jcap{\jref{JCAP}}

\end{document}